\documentstyle[12pt]{article}
\newcommand{\be}{\begin{equation}}
\newcommand{\ee}{\end{equation}}
\newcommand{\er}{\end{eqnarray}}
\newcommand{\br}{\begin{eqnarray}}
\newcommand{\dslash}{\partial\!\!\!/}
\newcommand{\aslash}{A\!\!\!/}
\newcommand{\bslash}{B\!\!\!/}
\newcommand{\kslash}{\kappa\!\!\!/}
\newcommand{\fslash}{f\!\!\!/}
\newcommand{\Dslash}{D\!\!\!\!/}
\begin{document}
\thispagestyle{empty}
$\phantom{x}$\vskip 0.618cm\par
{\huge \begin{center}Bosonisation and Duality Symmetry in the
Soldering Formalism
\end{center}}\par

\begin{center}
$\phantom{X}$\\
{\Large R.Banerjee\footnote{Present address: S.N.Bose National Centre for Basic
Sciences, Calcutta, India. email:rabin@boson.bose.res.in}   and C.Wotzasek
\footnote{email:clovis@if.ufrj.br}}\\[3ex]
{\em Instituto de F\'\i sica\\
Universidade Federal do Rio de Janeiro\\
21945, Rio de Janeiro, Brazil\\}
\end{center}\par
\begin{abstract}

\noindent We develop a technique that solders the dual aspects of some
symmetry. Using this technique it is possible to combine two
theories with such symmetries to yield a new effective theory.
Some applications in two and three dimensional
bosonisation are discussed. In particular, it is shown that
two apparently
independent three dimensional massive Thirring models with same
coupling but opposite mass signatures, in the long wavelegth limit,
combine by the process of bosonisation and soldering
to yield an effective massive Maxwell theory. 
Similar features also hold for quantum electrodynamics in three dimensions.
We also provide a systematic derivation of duality symmetric
actions and show that the soldering mechanism leads to a master
action which is duality invariant under a bigger set of
symmetries than is usually envisaged. The concept of duality
swapping is introduced and its implications are analysed. The example of
electromagnetic duality is discussed in details.
\end{abstract}


\vfill
\newpage

\section{Introduction}
This paper is devoted to analyse certain features and applications of
the soldering mechanism which is a new technique to work with
dual manifestations of some symmetry. 
This technique, which is independent of dimensional considerations,
essentially comprises in lifting the gauging
of a global symmetry to its local version and exploits certain concepts
introduced in a different context by Stone\cite{S} and
us\cite{W, ABW}.
The analysis is intrinsically
quantal without having any classical analogue. This is easily explained
by the observation that a simple addition of two independent classical
Lagrangeans is a trivial operation without leading to anything meaningful
or significant.

A particularly interesting application of soldering is in the
context
of bosonisation.
Although bosonisation was initially developed and fully 
explored in the context of two dimensions\cite{AAR}, more recently it has been
extended to higher dimensions\cite{M,C,RB,RB1}.
The importance of bosonisation lies in
the fact that it includes quantum effects already at the classical level. 
Consequently, different aspects and manifestations of quantum phenomena
may be investigated directly, that would otherwise be highly nontrivial
in the fermionic language. Examples of such applications are the 
computation of the current algebra\cite{RB} and the study of screening or
confinement in gauge theories\cite{AB}. An unexplored aspect of
bosonisation is revealed in the following question: given two independent
fermionic models which can be bosonised separately, under what 
circumstances is it possible to represent them by one single effective theory? 
The answer lies in the symmetries of the problem. Two  distinct models
displaying dual aspects of some symmetry can be combined by the
simultaneous implementation of bosonisation and soldering to yield a
completely new theory. 

The basic notions and ideas are first introduced in section 2
in the context of two
dimensions where bosonisation is known to yield exact results. The 
starting point is to take two distinct chiral Lagrangeans with opposite
chiralities. Using their bosonised expressions, the soldering mechanism
fuses, in a precise way, the left and right chiralities. This leads to
a general Lagrangean in which the chiral symmetry no longer exists, but
it contains  two extra parameters manifesting the bosonisation ambiguities.
It is shown that different  parametrisations lead either to 
the gauge invariant Schwinger model or the Thirring
model. 

Whereas the two dimensional analysis lays the foundations, the subsequent
three dimensional discussion (section 3)
illuminates the full power and utility of
the present approach for yielding new equivalences through bosonisation. 
While the bosonisation in these dimensions is not
exact, nevertheless, for massive fermionic models in the large mass or,
equivalently, the long wavelength limit, well defined local expressions
are known to exist\cite{C,RB}. Interestingly, these expressions exhibit a self
or an anti self dual symmetry that is dictated by the signature of the fermion
mass,
thereby providing a novel testing ground for our ideas. Indeed, two distinct
massive Thirring models with opposite mass signatures
are soldered to yield a massive Maxwell theory. This result
is vindicated by a direct comparison of the current correlation functions
obtained before and after the soldering process. As another instructive
application, the fusion of two models describing quantum electrodynamics
in three dimensions is considered. Results similar to the corresponding
analysis for the massive Thirring models are obtained.

In section 4 we present a systematic derivation of
electromagnetic duality
symmetric actions by converting the Maxwell action 
from a second order to a first
order form followed by a suitable relabelling of varibles which
naturally introduces an internal index. 
It is crucial to note that 
there are two distinct classes of
relabelling characterised by the opposite signatures of the
determinant of the $2\times 2$ orthogonal matrix defined in the internal 
space. 
Correspondingly, in this  space there are two  actions that are 
manifestly duality symmetric. 
Interestingly, their equations of motion are just the self and anti-self
dual solutions, where the dual field is defined in the internal space.
It
is also found that in all cases there is one (conventional duality) 
symmetry transformation which preserves the invariance of these actions
but there is another transformation which swaps the actions. We refer
to this property as swapping duality. This
indicates the possibility, in any dimensions, 
of combining the two actions to a master action
that would contain all the duality symmetries.
Indeed this construction is explicitly done by
exploiting the ideas of soldering.
The soldered master action also  has manifest Lorentz
or general coordinate invariance. The generators of the symmetry
transformations are also obtained. An appendix is included to
show briefly how these ideas can be carried over to two dimensions.

Section 5 contains our concluding remarks and observations.

\bigskip

\section{Bosonisation and soldering in two dimensions}

\bigskip

In this section we develop the ideas in the context of  two dimensions.
Consider, in particular, the following Lagrangeans with opposite chiralities, 
\begin{eqnarray}
{\cal L}_+&=&\bar\psi(i \dslash + e \aslash P_+)\psi\nonumber\\
{\cal L}_-&=&\bar\psi(i \dslash + e \aslash P_-)\psi
\label{10}
\end{eqnarray}
where $P_\pm$ are the projection operators,
\begin{equation}
P_\pm=\frac{1 \pm \gamma_5}{2}
\label{20}
\end{equation}
It is well known that the computation of the fermion determinant, which
effectively yields the bosonised expressions, is plagued by regularisation
ambiguities since chiral gauge symmetry cannot be preserved\cite{RJ}. Indeed an
explicit one loop calculation following Schwinger's point splitting method
\cite{RB2} yields the following bosonised expressions for the respective 
actions,
\begin{eqnarray}
\label{30}
W_+[\varphi] &=& \int d^2x \bar\psi (i\dslash+e\aslash_+)\psi= 
{1\over{4\pi}}\int d^2x\,\left(\partial_+
\varphi\partial_-\varphi +2 \, e\,A_+\partial_-\varphi + a\, 
e^2\, A_+ A_-\right)\nonumber\\
W_-[\rho]&=& \int d^2x \bar\psi(i\dslash+e\aslash_-)\psi= 
{1\over{4\pi}}\int d^2x\,\left(\partial_+\rho\partial_-
\rho +2 \,e\, A_-\partial_+\rho
+ b\, e^2\, A_+ A_-\right)
\end{eqnarray}
where the light cone metric has been invoked for convenience,
\begin{equation}
\label{35}
A_\pm = {1\over\sqrt 2}(A_0\pm A_1)=A^\mp \;\;\; ;\;\;\; \partial_\pm=
{1\over\sqrt 2}(\partial_0\pm \partial_1)=\partial^\mp
\end{equation}
Note that the regularisation or bosonisation ambiguity is manifested through
the arbitrary parameters $a$ and $b$. The latter ambiguity, in particular,
is clearly understood
since by using the normal bosonisation dictionary
$\bar\psi i\dslash\psi \rightarrow \partial_+\varphi\partial_-\varphi$
and $\bar\psi\gamma_\mu\psi\rightarrow{1\over\sqrt \pi}\epsilon_{\mu\nu}
\partial^\nu\varphi$ (which holds only for a gauge invariant theory),
the above expressions with $a=b=0$ are easily reproduced from (\ref{10}).

It is crucial to observe that different scalar fields $\varphi$ and $\rho$
have been used in the bosonised forms to emphasize the fact that the
fermionic fields occurring in the chiral components are uncorrelated.
It is the soldering process which will abstract a meaningful combination
of these components\cite{ABW}. This process essentially consists in lifting the
gauging of a global symmetry to its local version. Consider, therefore,
the gauging of the following global symmetry,
\begin{eqnarray}
\label{40}
\delta \varphi &=& \delta\rho=\alpha\nonumber\\
\delta A_{\pm}&=& 0
\end{eqnarray}

\noindent The variations in the effective actions  (\ref{30}) are found to be,

\begin{eqnarray}
\label{50}
\delta W_+[\varphi] &=& \int d^2x\, \partial_-\alpha \;J_+
(\varphi)\nonumber\\
\delta W_-[\rho]&=& \int d^2x\, \partial_+\alpha \;J_-(\rho)
\end{eqnarray}

\noindent  where the currents are defined as,

\begin{equation}
\label{60}
J_\pm(\eta)={1\over{2\pi}}(\partial_\pm\eta +\, e\,A_\pm)\;\;\; ; \;\;\eta=
\varphi , \rho
\end{equation}

\noindent  The important step now is to introduce the 
soldering field $B_\pm$ coupled with the currents so that,

\begin{equation}
\label{70}
W_\pm^{(1)}[\eta] = W_\pm[\eta] -\int d^2x\, B_\mp\, J_\pm(\eta)
\end{equation}

\noindent Then it is possible to define a modified action,

\begin{equation}
\label{80}
W[\varphi,\rho]= W_+^{(1)}[\varphi] + W_-^{(1)}[\rho]
 + {1\over{2\pi}} \int d^2x \, B_+ \,B_-
\end{equation}

\noindent which is invariant under an extended set of transformations that 
includes (\ref{40}) together with,

\begin{equation}
\delta B_{\pm}= \partial_{\pm}\alpha
\label{90}
\end{equation}

\noindent  Observe that the soldering field transforms exactly as a potential.
It has served its purpose of fusing the two chiral components. Since it 
is an auxiliary field, it can be eliminated from the invariant action 
(\ref{80}) by using the equations of motion. This will naturally solder the
otherwise independent chiral components and justifies its name as a soldering
field. The relevant solution is found to be,

\begin{equation}
\label{100}
B_\pm= 2\pi J_\pm
\end{equation}

\noindent Inserting this solution in (\ref{80}), we obtain,

\begin{equation}
\label{110}
W[\Phi]={1\over {4\pi}}\int d^2x\:\Big{\{}\Big{(}\partial_+
\Phi\partial_-\Phi + 2\,e\, A_+\partial_-\Phi - 2\,e\, A_-
\partial_+\Phi\Big{)} +(a+b-2)\,e^2\,A_+\,A_-\Big{\}}
\end{equation}

\noindent where,

\begin{equation}
\label{120}
\Phi=\varphi - \rho
\end{equation}

\noindent As announced, the action is no longer expressed in terms of the
different scalars $\varphi$ and $\rho$, but only on their  difference.
This difference is gauge invariant. 

Let us digress on the significance of the findings. At the classical fermionic
version, the chiral Lagrangeans are completely independent. Bosonising them
includes quantum effects, but still there is no correlation. The soldering
mechanism exploits the symmetries of the independent actions to
precisely combine them to yield a single action.
Note that the soldering works with the bosonised expressions. Thus the soldered
action obtained in this fashion corresponds to the quantum theory.    

We now show that different choices for the parameters $a$ and $b$
 lead to well known models.  To do this consider the variation of
(\ref{110}) under the conventional gauge transformations,
$\delta\varphi=\delta\rho=\alpha$ and $\delta A_\pm = \partial_\pm\alpha$.
It is easy to see that the expression
in parenthesis is gauge invariant. Consequently
a gauge invariant
structure for $W$ is obtained provided,
\begin{equation}
\label{130}
a+b-2=0
\end{equation}

The effect of soldering, therefore, has been to induce a lift of the initial
global symmetry (\ref{40}) to its local form. By functionally integrating
out the $\Phi$ field from (\ref{110}), we obtain,
\begin{equation}
\label{140}
W[A_+,A_-]=  -{ e^2\over 4\pi} \int d^2x\: \{A_+ 
{\partial_-\over \partial_+}A_+ 
+ A_- {\partial_+\over \partial_-}A_- - 2 A_+ A_-\}
\end{equation} 
which is the familiar structure for the gauge invariant action expressed in
terms of the potentials. The opposite chiralities of the 
two independent fermionic theories have been soldered to yield a gauge 
invariant action.

Some interesting observations are possible concerning the regularisation
ambiguity manifested by the parameters $a$ and $b$. As shown by
us \cite{ABW}, it is possible to uniquely fix these parameters
by demanding Bose symmetry \cite{RB3}. In the present case,
this symmetry corresponds to the left-right (or + -) symmetry in (\ref{30}),
thereby requiring $a=b$. Together with the condition (\ref{130}) this implies
$a=b=1$. This parametrisation has important consequences if a Maxwell term
was included from the beginning to impart dynamics. Then the soldering takes
place among two chiral Schwinger models\cite{JR} having opposite chiralities to
reproduce the usual Schwinger model\cite{JS}. 

Naively it may appear that the soldering of the left and right chiralities
to obtain a gauge invariant result is a
simple issue since adding the classical Lagrangeans
$\bar\psi\Dslash_+\psi$ and $\bar\psi\Dslash_-\psi$, with identical
fermion species, just yields the
usual vector Lagrangean $\bar\psi\Dslash\psi$. The quantum considerations 
are, however, much involved. The chiral determinants, as they occur,
cannot be even defined
since the kernels map from one chirality to the other so that there is no
well defined eigenvalue problem\cite{AW,RB3}. This is circumvented by
working with
$\bar\psi(i\dslash + e\aslash_{\pm})\psi$, that satisfy an eigenvalue
equation, from which their determinants may be computed. But now a simple
addition of the classical Lagrangeans does not reproduce the expected
gauge invariant form. At this juncture, the soldering process becomes
important. It systematically combined the quantised (bosonised)
expressions for the opposite chiral components. Note that {\it different}
fermionic species were considered so that this soldering does not have
any classical analogue, and is strictly a quantum phenomenon. This will 
become more transparent when the three dimensional case is discussed.

Next, we show how a different choice for the parameters $a$
and $b$ in (\ref{110}) leads to the Thirring model. Indeed it is precisely
when the mass term exists ($i.e., \,\, a+b-2\neq 0$), that (\ref{110})
represents
the Thirring  model. Consequently, this parametrisation complements that used
previously to obtain the vector gauge invariant structure. It is now easy to
see that the term in parentheses in (\ref{110}) corresponds to $\bar\psi
(i\dslash +e\aslash) \psi$ so that integrating out the nondynamical
 $A_\mu$ field
yields,
\be
{\cal L}=\bar\psi i\dslash\psi - \frac{g}{2}(\bar\psi\gamma_\mu\psi)^2\,\,\,\,
;\, g=\frac{4\pi}{a+b-2}
\label{A1}
\ee
which is just the Lagrangean for the usual Thirring model. It is known
\cite{SC}that 
this model is meaningful provided the coupling parameter satisfies the 
condition $g>-\pi$, so that,
\be
\label{A2}
\mid a+b \mid >2
\ee
This condition is the analogue of (\ref{130}) found earlier. As usual, there
is a one parameter arbitrariness. Imposing Bose symmetry implies that both
$a$ and $b$ are equal and lie in the range
\be
\label{A3}
1<\mid a\mid =\mid b\mid
\ee
This may be compared with the previous case where $a=b=1$ was necessary for
getting the gauge invariant structure.  Interestingly, the positive range
for the parameters in (\ref{A3}) just commences from this value.

Having developed and exploited the concepts of soldering in two dimensions,
it is natural to investigate their consequences in three dimensions. The
discerning reader may have noticed that it is essential to have dual
aspects of a symmetry that can be soldered to yield new information. In the
two dimensional case, this was the left and right chirality. Interestingly,
in three dimensions also, we have a similar phenomenon.

\bigskip

\section{Bosonisation and Soldering in three dimensions}

\bigskip

This section is devoted to an analysis of the soldering process in the massive
Thirring model and quantum electrodynamics in three dimensions.  
We first show that two apparently
independent massive Thirring models in the long wavelength limit combine,
at the quantum level, into a massive Maxwell theory. 
This is further vindicated by a direct comparison of the current correlation
functions following from the bosonization identities. 
These findings are also extended to include three dimensional
quantum electrodynamics. 
The new results and interpretations illuminate a close
parallel with the two dimensional discussion.

\bigskip

\subsection{The massive Thirring model}

\bigskip

In order to effect the soldering, the first step is to consider the
bosonisation of the massive Thirring
model in three dimensions\cite{C,RB}. This is therefore reviewed briefly. The
relevant current correlator generating functional,
in the Minkowski metric, is given by,
\be
\label{160}
Z[\kappa]=\int D\psi D\bar\psi \exp\Bigg(i\int d^3x\Bigg
[\bar\psi(i \dslash + m )\psi -\frac{\lambda^2}{2}
j_\mu j^\mu + \lambda j_\mu \kappa^\mu\Bigg]\Bigg)
\ee
where $j_\mu=\bar\psi\gamma_\mu\psi$ is the fermionic current. As usual,
the four fermion interaction can be eliminated by introducing an auxiliary
field,
\be
\label{170}
Z[\kappa]=\int D\psi D\bar\psi Df_\mu\exp\Bigg(i\int d^3x\Bigg
[\bar\psi\left(i \dslash + m +\lambda (\fslash +\kslash)\right)
\psi +\frac{1}{2} f_\mu f^\mu\Bigg]\Bigg)
\ee
Contrary to the two dimensional models, the fermion integration cannot be
done exactly. Under certain limiting conditions, however, this integration
is possible leading to closed expressions. A particularly effective choice
is the large mass limit in which case the fermion determinant yields a local
form. Incidentally, any other value of the mass leads to a nonlocal structure
\cite{RB1}.
The large mass limit is therefore very special. The leading term in this
limit was calculated by various means \cite{DJT} 
and shown to yield the Chern-Simons
three form. Thus the generating functional for the massive Thirring model in
the large mass limit is given by,
\be 
\label{180}
Z[\kappa]=\int Df_\mu 
\exp\Bigg( i\int d^3x\:\Bigg({\lambda^2\over{8\pi}}{m\over{\mid m\mid}}
\epsilon_{\mu\nu\lambda}f^\mu\partial^\nu f^\lambda +
\frac{1}{2} f_\mu f^\mu +\frac{\lambda^2}{4\pi}\frac{m}{\mid m\mid}
\epsilon_{\mu\nu\sigma}\kappa^\mu\partial^\nu f^\sigma\Bigg)\Bigg)
\ee
where the signature of the topological terms is dictated by the corresponding
signature of the fermionic mass term. 
In obtaining the above result a local counter term has been ignored. 
Such terms manifest the ambiguity in defining the time ordered product
to compute the correlation functions\cite{BRR}.
The Lagrangean in the above partition
function defines a self dual model introduced earlier \cite{TPN}. The massive
Thirring model, in the relevant limit, therefore bosonises to a self dual
model. It is useful to clarify the meaning of this self duality. The 
equation of motion in the absence of sources is given by,
\be
\label{190}
f_\mu =-{\lambda^2\over{4\pi}}{m\over{\mid m\mid}}
\epsilon_{\mu\nu\lambda}\partial^\nu f^\lambda  
\ee
from which the following relations may be easily verified,
\br
\label{200}
\partial_\mu f^\mu &=& 0\nonumber\\
\left(\Box + M^2\right)f_\mu &=& 0 \,\,\,\,\,\, ;\,\, 
M=\frac{4\pi}{\lambda^2}
\er
A field dual to $f_\mu$ is defined as,
\be
\label{210}
\tilde f_\mu = {1\over M} \epsilon_{\mu\nu\lambda}\partial^\nu f^\lambda
\ee
where the mass parameter $M$ is inserted for dimensional reasons. Repeating
the dual operation, we find,
\be
\label{220}
\tilde{\left(\tilde{f_\mu}\right)}= 
{1\over M} \epsilon_{\mu\nu\lambda}\partial^\nu\tilde{f^\lambda}=f_\mu
\ee
obtained by exploiting (\ref{200}), thereby validating the definition of
the dual field.  Combining these results with (\ref{190}),
we conclude that,
\be
\label{230}
f_\mu=- \frac{m}{\mid m \mid} \tilde f_\mu
\ee
Hence, depending on the sign of the fermion mass term, the bosonic theory
corresponds to a self-dual or an anti self-dual model.  Likewise, the Thirring
current bosonises to the topological current

\be
\label{235}
j_\mu = \frac{\lambda}{4\pi}\frac{m}{\mid m\mid}\epsilon_{\mu\nu\rho}
\partial^\nu f^\rho
\ee

The close connection with the two dimensional analysis is now evident.
There the starting point was to consider two distinct fermionic theories with 
opposite chiralities. In the present instance, the analogous thing is to
take two independent Thirring models with identical coupling strengths but
opposite mass signatures,
\br
\label{240}
{\cal L_+}&=&\bar\psi\left(i\dslash + m\right)\psi -\frac{\lambda^2}{2}
\left(\bar\psi\gamma_\mu\psi\right)^2\nonumber\\
{\cal L_-} &=& \bar \xi\left(i\dslash - m'\right)\xi - \frac{\lambda^2}{2} 
\left(\bar\xi\gamma_\mu\xi\right)^2
\er
Then the bosonised Lagrangeans are, respectively,
\br
\label{250}
{\cal L_+}&=&\frac{1}{2M} 
\epsilon_{\mu\nu\lambda}f^\mu\partial^\nu f^\lambda +
{1\over 2} f_\mu f^\mu\nonumber\\
{\cal L_-} &=&- \frac{1}{2M}
\epsilon_{\mu\nu\lambda}g^\mu\partial^\nu g^\lambda +
{1\over 2} g_\mu g^\mu
\er
where $f_\mu$ and $g_\mu$ are the distinct bosonic vector fields.  The current
bosonization formulae in the two cases are given by

\br
\label{255}
j_\mu^+ &=& \bar\psi\gamma_\mu\psi=\frac{\lambda}{4\pi} 
\epsilon_{\mu\nu\rho}\partial^\nu f^\rho\nonumber\\
j_\mu^- &=&\bar\xi\gamma_\mu\xi= - \frac{\lambda}{4\pi} 
\epsilon_{\mu\nu\rho}\partial^\nu g^\rho
\er

The stage is now set for soldering. Taking a cue from the two dimensional
analysis, let us consider the gauging of the following symmetry,
\be
\label{260}
\delta f_\mu = \delta g_\mu = 
\epsilon_{\mu\rho\sigma}\partial^\rho \alpha^\sigma
\ee
Under such transformations, the bosonised Lagrangeans change as,
\be
\label{270}
\delta{\cal L_\pm} = J_\pm^{\rho\sigma}(h_\mu)
 \partial_\rho\alpha_\sigma \,\,\,\,\, ;\,\, h_\mu=f_\mu,\,\,g_\mu
\ee
where the antisymmetric currents are defined by,
\be
\label{280}
J_\pm^{\rho\sigma}(h_\mu)= \epsilon^{\mu\rho\sigma}h_\mu \pm {1\over M}
\epsilon^{\gamma\rho\sigma}\epsilon_{\mu\nu\gamma}\partial^\mu 
h^\nu
\ee
It is worthwhile to mention that any other variation of the fields
(like $\delta{f_\mu}=\alpha_\mu$)is inappropriate because changes in
the two terms of the Lagrangeans cannot be combined to give a single
structure like (\ref{280}). We now introduce the soldering field coupled
with the antisymmetric currents. In
the two dimensional case this was a vector. Its natural extension now
is the antisymmetric second rank Kalb-Ramond tensor field $B_{\rho\sigma}$,
transforming in the usual way,
\be
\label{290}
\delta B_{\rho\sigma}=\partial_\rho\alpha_\sigma -
\partial_\sigma\alpha_\rho
\ee
Then it is easy to see that the modified Lagrangeans,
\be
\label{300}
{\cal L}_\pm^{(1)}={\cal L}_\pm - {1\over 2} J_\pm^{\rho\sigma}(h_\mu)
B_{\rho\sigma}
\ee
transform as,
\be
\label{310}
\delta{\cal L}_\pm^{(1)}=- {1\over 2} \delta J_\pm^{\rho\sigma}
B_{\rho\sigma}
\ee
The final modification consists in adding a term to ensure gauge invariance
of the soldered Lagrangean. This is achieved by,
\be
\label{320}
{\cal L}_\pm^{(2)}={\cal L}_\pm^{(1)} + {1\over 4} 
B^{\rho\sigma}B_{\rho\sigma}
\ee
A straightforward algebra shows that the following combination,
\br
\label{330}
{\cal L}_S &=& {\cal L}_+^{(2)}+{\cal L}_-^{(2)}\nonumber\\
&=&{\cal L}_+ + {\cal L}_- -{1\over 2}B^{\rho\sigma}
\left(J^+_{\rho\sigma}(f) + J^-_{\rho\sigma}(g)\right)
+{1\over 2} B^{\rho\sigma}B_{\rho\sigma}
\er
is invariant under the gauge transformations (\ref{260}) and (\ref{290}). 
The gauging of the symmetry
is therefore complete. To return to a description in terms of the original
variables, the auxiliary soldering field is eliminated from (\ref{330}) by 
using the equation of motion,
\be
\label{340}
B_{\rho\sigma}= {1\over 2} \left(J_{\rho\sigma}^+(f)+
J_{\rho\sigma}^-(g)\right)
\ee
Inserting this solution in (\ref{330}), the final soldered
Lagrangean is expressed
solely in terms of the currents involving the original fields,
\be
\label{350}
{\cal L}_S ={\cal L}_+ + {\cal L}_- -
{1\over 8}\left(J_{\rho\sigma}^+(f)+
J_{\rho\sigma}^-(g)\right)\left(J^{\rho\sigma}_+(f)+
J^{\rho\sigma}_-(g)\right)
\ee
It is now crucial to note that, by using the explicit structures for the
currents, the above Lagrangean is no longer a function of $f_\mu$ and $g_\mu$
separately, but only on the combination,
\be
\label{360}
A_\mu = {1\over{\sqrt{2} M}}\left(g_\mu - f_\mu\right)
\ee
with,
\be
\label{370}
{\cal L}_S = - \frac{1}{4} F_{\mu\nu}F^{\mu\nu} + 
{M^2\over 2}A_\mu A^\mu
\ee
where,
\be
\label{380}
F_{\mu\nu}= \partial_\mu A_\nu -\partial_\nu A_\mu
\ee
is the usual field tensor expressed in terms of the basic entity $A_\mu$.
Our goal has been achieved. The soldering mechanism has precisely fused
the self and anti self dual symmetries to yield a massive Maxwell theory
which, naturally, lacks this symmetry.

It is now instructive to understand this result by comparing the current
correlation functions.  The Thirring currents in the two models bosonise
to the topological currents (\ref{255}) in the dual formulation.  From a
knowledge of the field correlators in the latter case, it is therefore
possible to obtain the Thirring current correlators.  The field
correlators are obtained from the inverse of the kernels occurring in
(\ref{250}),

\br
\label{381}
\langle f_\mu(+k)\: f_\nu(-k)\rangle &=&
\frac{M^2}{M^2 - k^2}\left(i g_{\mu\nu} +
{1\over M}\epsilon_{\mu\rho\nu} k^\rho -
\frac{i}{M^2} k_\mu\:k_\nu\right)\nonumber\\
\langle g_\mu(+k)\: g_\nu(-k)\rangle &=&
\frac{M^2}{M^2 - k^2}\left(i g_{\mu\nu} -
{1\over M}\epsilon_{\mu\rho\nu} k^\rho -
\frac{i}{M^2} k_\mu\:k_\nu\right)
\er

\noindent where the expressions are given in the momentum space.
Using these in (\ref{255}), the current
correlators are obtained, 

\br
\label{382}
\langle j_\mu^+(+k) j_\nu^+(-k)\rangle &=&
\frac{M}{4\pi (M^2 - k^2)}
\left(i k^2 g_{\mu\nu}- ik_\mu\: k_\nu
+{1\over M}\epsilon_{\mu\nu\rho}
k^\rho\:k^2\right)\nonumber\\
\langle j_\mu^-(+k) j_\nu^-(-k)\rangle &=&
\frac{M}{4\pi (M^2 - k^2)}
\left(i k^2 g_{\mu\nu}- ik_\mu\: k_\nu
-{1\over M}\epsilon_{\mu\nu\rho}
k^\rho\:k^2\right)
\er
It is now feasible to construct a total current,

\be
\label{383}
j_\mu=j_\mu^+ + j_\mu^- = \frac{\lambda}{4\pi}
\epsilon_{\mu\nu\rho}\partial^\nu 
\left(f^\rho - g^\rho\right)
\ee
Then the correlation function for this current, in the original self dual
formulation, follows
from (\ref{382}) and noting that $\langle j_\mu^+\:
j_\nu^-\rangle =0$, which is a consequence of the
independence of $f_\mu$ and $g_\nu$;

\be
\label{384}
\langle j_\mu(+k)\: j_\nu(-k)\rangle = \langle j_\mu^+\: j_\nu^+\rangle +
\langle j_\mu^-\: j_\nu^-\rangle =
\frac{iM}{2\pi(M^2 -k^2)}
\left(k^2\: g_{\mu\nu} - k_\mu\: k_\nu\right)
\ee
The above equation is easily reproduced from the 
effective theory.  Using (\ref{360}), it is observed that the
bosonization of the composite current (\ref{383}) is
defined in terms of the massive vector field $A_\mu$,

\be
\label{385}
j_\mu=\bar\psi\gamma_\mu\psi +\bar\xi\gamma_\mu\xi =
-\sqrt{{M\over 2\pi}}\epsilon_{\mu\nu\rho}
\partial^\nu A^\rho
\ee
The current correlator is now obtained from the field
correlator $\langle A_\mu\: A_\nu\rangle$ given by the
inverse of the kernel appearing in (\ref{370}),

\be
\label{386}
\langle A_\mu(+k)\: A_\nu(-k)\rangle =
\frac{i}{M^2 - k^2}\left(g_{\mu\nu} - 
\frac{k_\mu\: k_\nu}{M^2}\right)
\ee
>From (\ref{385}) and (\ref{386}) the two point function
(\ref{384}) is reproduced, including the normalization.

We conclude, therefore, that the process of bosonisation and soldering of
 two massive Thirring models with opposite 
mass signatures, in the long wavelength limit,
 yields a massive
Maxwell theory. The bosonization of the composite current, obtained
by adding the separate contributions from the two models, is given in
terms of a topological current(\ref{385}) of the massive vector theory.
These are completely new results which cannot be obtained by a
straightforward application of conventional bosonisation techniques.
The massive modes in the original Thirring models are 
manifested in the two modes of (\ref{370}) so that there is a proper
matching in the degrees of freedom.
Once again it is reminded that the 
fermion fields for the models are different so that the analysis has no
classical analogue. Indeed if one considered the same fermion species,
then a simple addition of the classical Lagrangeans would lead to a 
Thirring model with a mass given by $m-m'$. 
In particular, this difference can be zero.
The bosonised version of such a massless model is
known \cite{M, RB1} to yield a highly nonlocal theory which has no connection
with (\ref{370}). Classically, therefore, there is no possibility of even
understanding, much less, reproducing the effective quantum result. 
In this sense the application in three dimensions
is more dramatic than the corresponding case of two dimensions.
        
\bigskip

\subsection{Quantum electrodynamics}

\bigskip

An interesting theory in which the preceding ideas may be
implemented is quantum electrodynmics, whose current
correlator generating
functional in an arbitrary covariant gauge is given by,
\be
\label{390}
Z[\kappa]=\int D\bar\psi\: D\psi\: DA_\mu\, \exp
\left\{i\int d^3x\:\left(\bar\psi\left(i\dslash + m +e\aslash
\right)\psi -{1\over 4} F_{\mu\nu}F^{\mu\nu} 
+{\eta\over 2}(\partial_\mu A^\mu)^2 + e j_\mu\kappa^\mu
\right)\right\}
\ee
where $\eta$ is the gauge fixing parameter and
$j_\mu = \bar\psi\gamma_\mu\psi$ is the current. 
As before, a one loop computation of the fermionic
determinant in the large mass limit yields,
\br
\label{400}
Z[\kappa]&=&\int  DA_\mu\, 
\exp { \lbrace} i\int d^3x\:\lbrack\frac{e^2}{8\pi}
\frac{m}{\mid m\mid}
\epsilon_{\mu\nu\lambda}A^\mu\partial^\nu A^\lambda
-{1\over 4} F_{\mu\nu}F^{\mu\nu}\nonumber\\
&+& \frac{e^2}{4\pi}
\frac{m}{\mid m \mid} \epsilon_{\mu\nu\rho}
\kappa^\mu\partial^\nu\: A^\rho +{\eta\over 2}
(\partial_\mu A^\mu)^2\rbrack{\rbrace}
\er
In the absence of sources, this just corresponds to the
topologically massive Maxwell-Chern-Simons theory, with
the signature of the topological term determined from that of
the fermion mass term. The equation of motion,

\be
\label{405}
\partial^\nu\, F_{\nu\mu} +\frac{e^2}{4\pi}
\frac{m}{\mid m \mid} \epsilon_{\mu\nu\lambda}
\partial^\nu A^\lambda = 0
\ee
expressed in terms of the dual tensor,

\be
\label{410}
F_\mu = \epsilon_{\mu\nu\lambda}
\partial^\nu A^\lambda
\ee
reveals the self (or anti self) dual property,
\be
\label{420}
F_\mu =\frac{4\pi}{e^2}\frac{m}{\mid m\mid}
\epsilon_{\mu\nu\lambda}\partial^\nu F^\lambda
\ee
which is the analogue of (\ref{190}).  In this fashion the
Maxwell-Chern-Simons theory
manifests the well known \cite{DJ, BRR, BR} mapping with the self
dual models considered in the previous subsection.
The difference is that the self duality in the former,
in contrast to the latter, is contained in the dual
field (\ref{410}) rather than in the basic field defining
the theory. This requires some modifications in the
ensuing analysis.  Furthermore, the bosonization of the
fermionic current is now given by the topological current
in the Maxwell-Chern-Simons theory,

\be
\label{425}
j_\mu = \frac{e}{4\pi} \frac{m}{\mid m \mid}
\epsilon_{\mu\nu\lambda}\partial^\nu A^\lambda
\ee

Consider, therefore, two independent models describing quantum 
electrodynamics with opposite signatures in the mass terms,

\br
\label{426}
{\cal L}_+ &=& \bar\psi\left( i\dslash +m
+e\aslash\right)\psi -{1\over 4} 
F_{\mu\nu}(A) F^{\mu\nu}(A)\nonumber\\
{\cal L}_- &=& \bar\xi\left( i\dslash -m'
+e\bslash\right)\xi -{1\over 4} 
F_{\mu\nu}(B) F^{\mu\nu}(B)
\er

whose bosonised versions in an appropriate limit are given by,
\br
\label{430}
{\cal L}_+ &=& -\frac{1}{4} F_{\mu\nu}(A)+\frac{M}{2}
\epsilon_{\mu\nu\lambda}A^\mu\partial^\nu A^\lambda 
\,\,\,\,\, ; \,\, M=\frac{e^2}{4\pi}\nonumber\\
{\cal L}_- &=& -\frac{1}{4} F_{\mu\nu}(B)-\frac{M}{2}
\epsilon_{\mu\nu\lambda}B^\mu\partial^\nu B^\lambda
\er
where $A_\mu$ and $B_\mu$ are the corresponding potentials. 
Likewise, the corresponding expressions for the bosonized
currents are found from the general structure (\ref{425}),

\br
\label{435}
j_\mu^+ &=&\bar\psi\gamma_\mu\psi= \frac{M}{e} \epsilon_{\mu\nu\lambda}
\partial^\nu A^\lambda\nonumber\\
j_\mu^- &=&\bar\xi\gamma_\mu\xi= -\frac{M}{e} \epsilon_{\mu\nu\lambda}
\partial^\nu B^\lambda
\er
To proceed with
the soldering of the above models, take  the symmetry transformation,
\be
\label{440}
\delta A_\mu=\alpha_\mu
\ee
Such a transformation is spelled out by recalling (\ref{260}) and the 
observation that now (\ref{410}) simulates the $f_\mu$ field in the previous
case. Under this variation, the Lagrangeans (\ref{430}) change as,
\be
\label{450}
\delta{\cal L}_\pm =J_\pm^{\rho\sigma}(P)\partial_\rho\alpha_\sigma
\,\,\,\,\, ; \,\, P=A,B
\ee
where the antisymmetric  currents are defined by,
\be
\label{460}
J_\pm^{\rho\sigma}(P)=\pm m \epsilon^{\rho\sigma\mu}P_\mu -F^{\rho\sigma}(P)
\ee
Proceeding as before, 
it is now straightforward to deduce the final Lagrangean that will be
gauge invariant. This is given by,
\be
\label{490}
{\cal L}_S = \tilde{\cal L}_+ + \tilde{\cal L}_-
\,\,\,\,\, ; \,\, \delta{\cal L}_S =0
\ee
where the  iterated pieces are,
\be
\label{500}
\tilde{\cal L}_\pm= {\cal L}_\pm -{1\over 2} J_\pm^{\rho\sigma}
B_{\rho\sigma} -{1\over 4}B_{\rho\sigma}B^{\rho\sigma}
\ee
To obtain the effective soldered Lagrangean, the auxiliary
$B_{\rho\sigma}$ field is eliminated 
by using the equation of motion and the final result is,
\be
\label{530}
{\cal L}_S =-{1\over 4} F_{\mu\nu}(G)F^{\mu\nu}(G) 
+\frac{M^2}{2}G_\mu G^\mu
\ee
written in terms of a single field,
\be
\label{540}
G_\mu =\frac{1}{\sqrt{2}}\left(A_\mu - B_\mu\right)
\ee
The Lagrangean (\ref{530}) governs the dynamics of a massive Maxwell theory.

As before, we now discuss the implications for the current
correlation functions.  These functions in the original models
describing electrodynamics can be obtained from the mapping
(\ref{435}) by exploiting the field
correlators found by inverting the kernels occurring in
(\ref{430}).  The results, in the momentum space, are

\br
\label{546}
\langle j_\mu^+(+k)\:j_\nu^+(-k)\rangle &=&
i\left(\frac{M}{e}\right)^2\frac{1}{M^2 -k^2}
\left[k^2 g_{\mu\nu} -k_\mu\: k_\nu -
i M\epsilon_{\mu\nu\rho}k^\rho\right]
\nonumber\\
\langle j_\mu^-(+k)\:j_\nu^-(-k)\rangle &=&
i\left(\frac{M}{e}\right)^2\frac{1}{M^2 -k^2}
\left[k^2 g_{\mu\nu} -k_\mu\: k_\nu +
i M\epsilon_{\mu\nu\rho}k^\rho\right]
\er
Next, defining a composite current,

\be
\label{547}
j_\mu = j_\mu^+ + j_\mu^- = \frac{M}{e} 
\epsilon_{\mu\nu\lambda}\partial^\nu
\left(A^\lambda - B^\lambda\right)
\ee
it is simple to obtain the relevant correlator
by exploiting the results for $j_\mu^+$ and $j_\mu^-$ from (\ref{546}),

\be
\label{548}
\langle j_\mu(+k)\: j_\nu(-k)\rangle =
2i\left(\frac{M}{e}\right)^2\frac{1}{M^2 -k^2} 
\left(k^2 g_{\mu\nu} - k_\mu\: k_\nu\right)
\ee
In the bosonized version obtained from the soldering,
(\ref{547}) represents the mapping,

\be
\label{549}
j_\mu=\bar\psi\gamma_\mu\psi +\bar\xi\gamma_\mu\xi =
\sqrt 2 {M\over e} \epsilon_{\mu\nu\lambda}
\partial^\nu G^\lambda
\ee
where $G_\mu$ is the massive vector field (\ref{540})
whose dynamics is governed by the Lagrangean (\ref{530}). 
In this effective description the result (\ref{548}) is
reproduced from (\ref{549}) by using the correlator
of $G_\mu$ obtained from (\ref{530}), which is exactly identical to
(\ref{386}).

Thus the combined effects of bosonisation and soldering  of two
independent
quantum electrodynamical models with appropriate mass signatures 
 yield a massive Maxwell theory. In the self dual version
the massive modes are the topological excitations in the Maxwell-Chern-Simons
theories. These are combined into the two usual massive modes in the effective
massive vector theory.
A complete correspondence among the composite current correlation
functions in the original models and in their dual bosonised
description was also established.  The comments made
in the concluding part of the last subsection naturally apply also in this
instance.

It is interesting to note from the above analysis  that
in the quadratic approximation in
the large mass limit, the massive Thirring model is equivalent
to quantum electrodynamics. Furthermore a direct comparison of
the mass terms in the effective vector theory reveals the dual
nature of this equivalence since the coupling in one case is
related to the inverse coupling in the other. In the next
section we analyse the consequences of soldering for
electromagnetic duality in four dimensions.

\bigskip

\section{Electromagnetic duality and soldering}

\bigskip

In recent years the old idea \cite{O, Z, DT} of electromagnetic duality
has been
revived with considerable attention and emphasis \cite{SS, GR,
NB, DGHT, AG}. 
Different directions of research \cite{SS, KP, PST, G} 
include an 
abstraction of manifestly covariant forms  for duality symmetric 
actions or an explicit proof of the equivalence of such actions 
with the original nonduality
symmetric actions.
In spite of this spate of papers there does not seem to be a
simple clear cut way of obtaining duality symmetric actions.
We first discuss such an approach introducing simultaneously the
concept of duality swapping. This is  followed  by analysing
the effects of soldering. To show the generality of the ideas
the case of a scalar field in two dimensions is also analysed in
an appendix.

Let us start with
the usual Maxwell Lagrangean,
\be
{\cal L}=-\frac{1}{4}F_{\mu\nu}F^{\mu\nu}
\label{m10}
\ee
which is expressed in terms of the electric and magnetic fields
as,\footnote{Bold face letters denote three vectors.}
\be
{\cal L}= \frac{1}{2}\Big(\bf E^2-\bf B^2\Big)
\label{m20}
\ee
where,
\br
E_i&=&-F_{0i}=-\partial_0 A_i+\partial_i A_0\nonumber\\
B_i&=&\epsilon_{ijk}\partial_j A_k
\label{m30}
\er
The following duality transformation,
\be
\bf E\rightarrow \mp\bf B\,\,\,\,;\,\,\,\,\bf B\rightarrow \pm \bf E
\label{m40}
\ee
is known to preserve the invariance of the full set comprising
Maxwell's equations and the Bianchi identities although the Lagrangean
changes its sign. To have a duality symmetric Lagrangean,
the primary step is to recast (\ref{m20})
in a symmetrised first order form by introducing an auxiliary field,
\be
{\cal L}=\frac{1}{2}\Big({\bf P}.\dot{\bf A}-\dot{\bf P}.{\bf A}\Big)
-\frac{1}{2}{\bf P}^2-\frac{1}{2}{\bf B}^2
+A_0\bf\nabla.\bf P
\label{m50}
\ee
A suitable change of
variables is now invoked which naturally introduces an internal index. 
Significantly, there are two possibilities
which translate
from the old set $(\bf P, \bf A)$ to the new ones $(\bf A_1, \bf A_2)$.
It is, however, important to note that the Maxwell theory has a constraint
that is implemented by the Lagrange multiplier $A_0$. The redefined variables
are chosen such that this constraint is automatically satisfied,
\br
\bf P&\rightarrow& \bf B_2\,\,\,\,;\,\,\,\,\bf A\rightarrow \bf A_1\nonumber\\
\bf P&\rightarrow& \bf B_1\,\,\,\,;\,\,\,\,\bf A\rightarrow \bf A_2
\label{m60}
\er
It is now simple to show that, in terms of the redefined variables, the
original Maxwell Lagrangean takes the form,
\be
{\cal L}_\pm={1\over 2}\left(\pm\bf {\dot A}_\alpha
\epsilon_{\alpha\beta}\bf B_\beta
-\bf B_\alpha\bf B_\alpha\right)
\label{m70}
\ee
where $\epsilon_{\alpha\beta}$ is the second rank antisymmetric
tensor defined in the internal space 
with $\epsilon_{12}=1$.
Adding a total derivative that would leave the equations of motion unchanged,
this Lagrangean is expressed directly in terms of the electric and magnetic
fields,
\be
{\cal L}_\pm={1\over 2}\left(\pm\bf B_\alpha
\epsilon_{\alpha\beta}\bf E_\beta
-\bf B_\alpha\bf B_\alpha\right)
\label{m70a}
\ee

Observe that only one of the
above structures (namely, ${\cal L}_-$) was given earlier in \cite{SS}.
The presence of two structures leads to some interesting consequences.
Let us first introduce the proper and improper rotation matrices
parametrised by the angles $\theta$ and $\varphi$, respectively,

\br
R^+(\theta)
=
\left(\begin{array}{cc}
{\cos\theta} & {\sin\theta} \\
{-\sin\theta} &{\cos\theta} \end{array}\right)
\label{matrix1}
\er
\br
R^-(\varphi)
=
\left(\begin{array}{cc}
{\sin\varphi} & {\cos\varphi} \\
{\cos\varphi} &{-\sin\varphi} \end{array}\right)
\label{matrix}
\er

Both the Lagrangeans are duality symmetric under the proper $(SO(2))$
transformations,
\br
{\bf{E}}_\alpha&\rightarrow& R^+_{\alpha\beta}{\bf{E}}_\beta\nonumber\\
{\bf{B}}_\alpha&\rightarrow& R^+_{\alpha\beta}{\bf{B}}_\beta
\er
\label{100a}

Interestingly, under the improper rotations,
\br
{\bf{E}}_\alpha&\rightarrow& R^-_{\alpha\beta}{\bf{E}}_\beta\nonumber\\
{\bf{B}}_\alpha&\rightarrow& R^-_{\alpha\beta}{\bf{B}}_\beta
\label{1000a}
\er
the Lagrangean ${\cal L}_+$ goes over to ${\cal L}_-$ and vice
versa. We refer to this property as swapping duality.
Note that the discretised version of the above equation is obtained by setting 
$\varphi =0$,
\br
{\bf{E}}_\alpha&\rightarrow& \sigma_1^{\alpha\beta}{\bf{E}}_\beta\nonumber\\
{\bf{B}}_\alpha&\rightarrow& \sigma_1^{\alpha\beta}{\bf{B}}_\beta
\label{110a}
\er
It is precisely the $\sigma_1$ matrix that reflects the proper into
improper rotations,
\be
\label{everton}
R^+(\theta) \sigma_1=R^-(\theta)
\ee
which illuminates the reason behind the swapping of the Lagrangeans in
this example.

The generators of the $SO(2)$ rotations are given by,
\be
\label{john}
Q^{(\pm)}=\mp\frac{1}{2}\int d^3x\,\, {\bf{A}}^\alpha\,.\,{\bf{B}}^\alpha
\ee
so that,
\be
\label{gt}
{\bf{A}}_\alpha\rightarrow {\bf{A}}'_\alpha=e^{-iQ\theta}{\bf {A}}_\alpha
e^{iQ\theta} 
\ee
This can be easily verified by using the basic brackets following
from the symplectic structure of the theory,
\be
\label{girotti}
\Big [A^i_\alpha(x),
\epsilon^{jkl}\partial^k A^l_\beta(y)\Big]=\pm i\delta^{ij}
\epsilon_{\alpha\beta} \delta({\bf{x}}-{\bf{y}})
\ee
It is useful to comment on the significance of the above analysis. Since
the duality symmetric Lagrangeans have been obtained directly from
the Maxwell Lagrangean, it is redundant to show the equivalence of
the former expressions with the latter, which is an essential
perquisite in other approaches. Furthermore, since classical
equations of motion have not been used at any stage, the purported
equivalence holds at the quantum level. The need for any explicit
demonstration of this fact, which has been the motivation of several
recent papers, becomes, in this analysis, superfluous. 
A related observation is that the usual way of showing the classical
equivalence is to use the
equations of motion to eliminate one component from (\ref{m70}),
thereby leading to the Maxwell Lagrangean in the temporal $A_0=0$ gauge.
This is not surprising since the change of variables leading from the
second to the first order form solved the Gauss law thereby
eliminating the multiplier. Finally, note that there are 
two distinct structures for the duality symmetric
Lagrangeans. These must correspond to the opposite aspects of some
symmetry, which is next unravelled. By looking at the equations of
motion obtained from (\ref{m70}),
\be
\bf {\dot A}_\alpha =
 \pm\epsilon_{\alpha\beta}\bf \nabla \times \bf A_\beta
\label{m80}
\ee 
it is possible to
verify that these are just the 
self and anti-self dual solutions,
\be
F_{\mu\nu}^\alpha=\pm\epsilon^{\alpha\beta}\mbox{}^* F_{\mu\nu}^\beta
\,\,;\,\,\mbox{}^* F_{\mu\nu}^\beta=\frac{1}{2}\epsilon_{\mu\nu\rho\lambda}
F^{\rho\lambda}_\beta
\label{m90}
\ee
obtained by setting $A_0^\alpha=0$.  As shown in the appendix,
in the two dimensional
theory the equation of motion naturally assumes a covariant structure.
Here, on the other hand, the introduction of $A_0^\alpha$ is necessary
since, as shown earlier, the term involving $A_0$  drops out in the
construction of the duality invariant Lagrangean. This is because $A_0$ is a
multiplier enforcing the Gauss constraint. Such a 
feature distinguishes a gauge theory from the non gauge theory discussed
in the two dimensional example.
Following our
general strategy, the next task is to solder the two Lagrangeans
(\ref{m70}). Consider then the gauging of the following symmetry,
\be
\delta \bf H_\alpha={\bf{h}}_\alpha \,\,\,;\,\,\,\bf H=\bf P, \bf Q
\label{m100}
\ee
where $\bf P$ and $\bf Q$ denote the basic fields in the Lagrangeans
${\cal L}_+$ and ${\cal L}_-$, respectively. The Lagrangeans
transform as,
\be
\delta {\cal L}_\pm=\epsilon_{\alpha\beta} 
\Big(\bf \nabla\times{\bf{h}}_\alpha\Big).\bf J_\beta^\pm
\label{m110}
\ee
with the currents defined by,
\be
\bf J_\alpha^\pm(\bf H)=\Big(\mp\bf \dot{H}_\alpha+\epsilon_{\alpha\beta}
\bf \nabla\times\bf H_\beta\Big)
\label{m120}
\ee
Next, the soldering  field $\bf W_\alpha$ is introduced which transforms as,
\be
\delta\bf W_\alpha =-\epsilon_{\alpha\beta}\bf\nabla\times{\bf{h}}_\beta
\label{m130}
\ee
Following standard steps as outlined previously, the final  Lagrangean
which is invariant under the complete set of transformations (\ref{m100})
and (\ref{m130}) is obtained,
\be
{\cal L}={\cal L}_+({\bf P})+{\cal L}_-({\bf Q})-{\bf W}^\alpha\,.\,\Big(\bf
J_\alpha^+ (\bf P)+ J_\alpha^- (\bf Q)\Big)- {\bf W}_\alpha^2
\label{m130a}
\ee
Eliminating the soldering field by using the equations of motion, the effective
soldered Lagrangean following from (\ref{m130a}) is derived,
\be
{\cal L}=\frac{1}{4}\Bigg(\bf {\dot G}_\alpha.{\dot G}_\alpha
-\bf\nabla\times\bf G_\alpha.\bf\nabla\times\bf G_\alpha\Bigg)
\label{m140}
\ee
where the composite field is given by the  combination,
\be
\bf G_\alpha=\bf P_\alpha-\bf Q_\alpha
\label{m150}
\ee
which is invariant under (\ref{m100}).
It is interesting to note that, reintroducing the $G_0^\alpha$ variable, 
this is nothing but the Maxwell Lagrangean
with a doublet of fields,
\be
{\cal L}=-\frac{1}{4} G_{\mu\nu}^\alpha G^{\mu\nu}_\alpha\,\,\,;\,\,\,
G_{\mu\nu}^\alpha=\partial_\mu  G_\nu^\alpha-
\partial_\nu  G_\mu^\alpha
\label{m160}
\ee
It is now possible to show that by reducing
(\ref{m160}) to a first order from, we exactly obtain the two types of
the duality symmetric Lagrangeans (\ref{m70a}). This shows the equivalence of
the soldering and reduction (i.e. conversion of a second order Lagrangean to
its first order form) processes.

In terms of the original $ P$ and $Q$ fields a generalised 
Polyakov-Weigmann like identity \cite{ABW} is obtained,

\br
{\cal L}(P-Q)&=&{\cal L}(P)+{\cal L}(Q)-2 W_{i, \alpha}^+( P)
W_{i, \alpha}^-( Q)\nonumber\\
W_{i, \alpha}^\pm(H) &=&\frac{1}{\sqrt 2}\Big(F_{0i}^\alpha(H)
\pm \epsilon_{ijk}\epsilon_{\alpha\beta}
F_{jk}^\beta(H)\Big)\,\,\,;\,\,\, H=P, Q
\label {m170}
\er
With respect to the gauge transformations (\ref{m100}), the above identity
shows that a contact term is necessary to restore the gauge invariant action
from two gauge variant forms. This, it may be recalled, is just the basic 
content of the Polyakov-Weigmann identity. It is interesting to note that
the ``mass" term appearing in the above identity is composed of parity
preserving pieces $W_{i, \alpha}^\pm$, thanks to the presence of the
compensating $\epsilon$-factor from the internal space.

A particularly illuminating way of rewriting the Lagrangean (\ref{m160}) is,
\br
{\cal L} &=&-\frac{1}{8}\Big( G_{\mu\nu}^\alpha + \epsilon^{\alpha\beta}
\mbox{}^* G_{\mu\nu}^\beta\Big)
\Big( G^{\mu\nu}_\alpha -\epsilon_{\alpha\rho}
\mbox{}^* G^{\mu\nu}_\rho\Big)\nonumber\\
&=&-\frac{1}{8}\Big( G_{\mu\nu}^\alpha + 
\tilde G_{\mu\nu}^\alpha\Big)
\Big( G^{\mu\nu}_\alpha -\tilde G^{\mu\nu}_\alpha\Big)
\label{m180}
\er
where, in the second line, the generalised Hodge dual $(\tilde
G)$ in the space 
containing the internal index has been defined in terms of the
usual Hodge dual $(\mbox{}^*G)$ 
to explicitly show the soldering of the self and anti self dual
solutions. 
The above Lagrangean manifestly displays the following duality symmetries,
\be
A_{\mu}^\alpha \rightarrow R_{\alpha\beta}^\pm A_{\mu}^{\beta}
\label{m190}
\ee

\noindent where, without any loss of generality, 
we may denote the composite field,
of which $G_{\mu\nu}$ is a function, by $A$. 
The generator of the $SO(2)$ rotations is now given by,
\be
\label{m191}
Q=\int d{\bf{x}}\,\,
\epsilon^{\alpha\beta}{\bf {\Pi}}^\alpha\,\,.\,\,{\bf {A}}^\beta
\ee

Now observe that the master Lagrangean was
obtained from the soldering of two distinct Lagrangeans (\ref{m70}). The
latter were duality symmetric under the proper rotations while
the improper ones effected a swapping. The soldered
Lagrangean is therefore duality symmetric under the complete set
of transformations 
(\ref{m190})
implying that it  contains a bigger set of duality
symmetries than (\ref{m70}). Significantly, it is also manifestly
Lorentz invariant. Furthermore, recall 
that under the transformations mapping the field
to its dual, the original 
Maxwell equations are invariant but the  Lagrangean changes its signature.
The corresponding transformation in the $SO(2)$ space is
given by, 
\be
\label{0}
G_{\mu\nu}^\alpha \rightarrow R^+_{\alpha\beta}
\mbox{}^*G_{\mu\nu}^\beta
\ee
which, written in component notation, looks like,
\be
\bf E^\alpha\rightarrow\mp\epsilon^{\alpha\beta}\bf B^\beta\,\,\,;\,\,\,
\bf B^\alpha\rightarrow\pm\epsilon^{\alpha\beta}\bf E^\beta
\label{m200}
\ee
The standard duality symmetric Lagrangean fails to manifest this property.
However, as may be easily checked, the equations of motion obtained from 
the master Lagrangean swap with the corresponding Bianchi identity
while the Lagrangean flips sign. In this manner the original property
of the second order Maxwell Lagrangean is retrieved.
Note furthermore that the master Lagrangean possesses the $\sigma_1$ symmetry
(which is just the discretised version of $ R^-$ with $\varphi=0$), 
a feature expected for two
dimensional theories. In the appendix, a 
similar phenomenon is reported
where the master action in two dimensions reveals the $SO(2)$
symmetry usually associated with four dimensional theories.

\subsection{Coupling to gravity}

The effects of coupling to gravity are straightforwardly
included by starting  from the following Lagrangean,

\be
{\cal
L}=-\frac{1}{4}\sqrt{-g}g^{\mu\alpha}g^{\nu\beta}F_{\mu\nu}F_{\alpha\beta}
\label{g10} 
\ee
>From our experience in the usual Maxwell theory we know that an eventual 
change of variables eliminates the Gauss law so that the term involving the
multiplier $A_0$ may be ignored from the outset. Expressing (\ref{g10}) in
terms of its components to separate explicitly the first and second order
terms, we find,
\be
\label{g20}
{\cal L}=\frac{1}{2}\dot A_i M^{ij} \dot A_j + M^i \dot A_i +M
\ee
where,
\br
M^{ij}&=&\sqrt{-g}\Big(g^{0i}g^{0j}- g^{ij}g^{00}\Big)\nonumber\\
M^{i}&=&\sqrt{-g} g^{0k}g^{ji} F_{jk}
\nonumber\\
M&=&\frac{1}{4}\sqrt{-g} g^{ij}g^{km} F_{im}F_{kj}
\label{g30}
\er
Now reducing the Lagrangean to its first order form, we obtain,
\be
\label{g40}
{\cal L}= P^i E_i-\frac{1}{2}P^i M_{ij} P^j -\frac{1}{2}M^i M_{ij} M^j
+P^i M_{ij} M^j +M
\ee
where $\dot A_i$ has been replaced by $E_i$ and 
$M_{ij}$ is the inverse of $M^{ij}$,
\be
\label{g50}
M_{ij}= \frac{-1}{\sqrt{-g}g^{00}} g_{ij}
\ee
with,
\be
\label{g60}
g^{\mu\nu}g_{\nu\lambda}=\delta^\mu_\lambda
\ee
Next, following the Maxwell example, 
introduce the standard change of variables which solves the Gauss
constraint,
\br
\label{g70}
E_i &\rightarrow & E_i^{(1)}\nonumber\\
P^i&\rightarrow &\pm B^{i(2)}
\er
the Lagrangean (\ref{g40}) is expressed in the desired form,
\br
\label{g80}
{\cal L}_\pm=&\pm & E_i^\alpha\epsilon^{\alpha\beta}B_\beta^i
+\frac{1}{\sqrt{-g}g^{00}}g_{ij} B^i_\alpha B_\alpha^j\nonumber\\
&\pm & \frac {g^{0k}}{g^{00}} \epsilon_{ijk}
\epsilon^{\alpha\beta}B_\alpha^i B_\beta^j
\er

Once again there are two duality symmetric actions corresponding to ${\cal
L}_\pm$. The enriched nature of the duality and swapping 
symmetries under a bigger set
of transformations, the constructing of a master Lagrangean from
soldering of ${\cal L}_+$ and ${\cal L}_-$, the corresponding
interpretations, all go through exactly as in the flat metric case.
Incidentally, the structure for ${\cal L}_-$ only was previously given in
\cite {SS}. 

\bigskip

\section{Conclusions}

The present analysis clearly revealed the possibility of obtaining new
results  by combining two apparently
independent theories into a single effective theory. 
The essential ingredient was that these theories
must possess the dual aspects of the same symmetry. Then, by a systematic
application of the soldering technique, it was feasible to abstract
a meaningful combination of such models, which can never be obtained
by a naive addition of the classical Lagrangeans. Detailed
applications of this technique were presented in the context of
bosonisation and duality symmetry. Simultaneously, a 
method for obtaining duality symmetric actions that were
soldered was also developed.

The basic notions and ideas were first  illustrated in 
two dimensional bosonisation. Bosonised expressions for distinct
chiral Lagrangeans were soldered to reproduce either the usual gauge
invariant theory or the Thirring model. Indeed, the soldering mechanism
that fused the opposite chiralities clarified several aspects of the  
ambiguities occurring in bosonising chiral Lagrangeans. It was  
shown that unless Bose symmetry was imposed as an additional restriction,
there is a whole one parameter class of bosonised solutions for the chiral
Lagrangeans that can be soldered to yield the vector gauge invariant result.
The close connection between Bose symmetry and gauge invariance was thereby
established, leading to a unique parametrisation. Similarly, using a 
different parametrisation, the soldering of the chiral Lagrangeans led to
the Thirring model. Once again there was a one parameter ambiguity unless
Bose symmetry was imposed. If that was done, there was a specified range of 
solutions for the chiral Lagrangeans that combined to yield a well defined
Thirring model.

The elaboration of our methods was done by considering the massive version
of the Thirring model and quantum electrodynamics in three dimensions.
By the process of bosonisation such models, in the long wavelength limit, 
were cast in a form which 
manifested a self dual symmetry. This was a basic prerequisite
for effecting
the soldering. It was explicitly shown that two distinct massive Thirring 
models, with opposite mass signatures, combined to a massive Maxwell theory.
The Thirring current correlation functions calculated either in the original
self dual formulation or in the effective massive vector theory yielded 
identical results, showing the consistency of our approach. The application
to quantum electrodynamics followed along similar lines.

The present work also revealed a unifying structure behind the construction of
the various duality symmetric actions ((76), (110), (128)). 
The essential ingredient was the
conversion of the second order action into a first order form followed by
an appropriate redefinition of variables such that these may be denoted
by  an internal index. The duality naturally occurred in this internal
space. Since the duality symmetric actions were directly derived from the
original action the proof of their equivalence becomes superfluous. This
is otherwise essential where such a derivation is lacking and recourse is
taken to either equations of motion  or some hamiltonian analysis.

A notable feature of the analyis was the revelation of a whole class of
new symmetries and their interrelations. Different aspects of this feature were
elaborated.  To be precise, it was shown that there are actually two
\footnote {Note that usual discussions of duality symmetry consider only
one of these actions, namely ${\cal L}_-$.}
duality symmetric actions $({\cal L}_\pm)$ 
for the same theory. These actions carry the opposite (self and anti self
dual) aspects of some symmetry and their occurrence  was
essentially tied to the fact that there were two distinct classes in which
the renaming of variables was possible, depending on the signature of the
determinant specifying the proper or improper rotations. 
To discuss further the implications of this pair of duality symmetric
actions it is best to compare with the existing results. This also serves
to put the present work in a proper perspective. It should be mentioned
that the analysis for two \footnote {see appendix}
and four dimensions were generic for $4k+2$ 
and $4k$ dimensions, respectively.

It is usually observed \cite{DGHT} 
that the invariance of the actions in different
$D$-dimensions is preserved by the following groups,
\be
\label{c1}
{\cal G}_d=Z_2 \,\,\,;\,\,\, D=4k+2
\ee
and,
\be
\label{c2}
{\cal G}_c= SO(2)\,\,\,;\,\,\, D=4k
\ee
which are called the ``duality groups". The $Z_2$ group is a discrete
group with two elements, the trivial identity and the $\sigma_1$ matrix.
Observe an important difference
since in one case this group is continous while in the other it is
discrete. In our exercise this was easily verified by the pair of duality
symmetric actions ${\cal L}_\pm$. The new ingredient was that nontrivial
elements of these
groups ($\sigma_1$ for $Z_2$ and $\epsilon$ for $SO(2)$)
were also responsible for the swapping ${\cal L}_+\leftrightarrow {\cal
L}_-$, but in dimensions different from where they act as elements of duality
groups. In other words the ``duality swapping matrices"
$\Sigma_s$ are given by,
\br
\label{c3}
{\Sigma}_s &=&\sigma_1\,\,\,;\,\,\, D=4k\nonumber\\
 &=& \epsilon\,\,\,;\,\,\, D=4k+2
\er
A comparison with (\ref{c1}) and (\ref{c2}) shows the reversal of roles of the 
matrices with regard to the dimensionality of space time.

It was next shown that ${\cal L}_\pm$ contained the self and anti-self
dual aspects of some symmetry. Consequently, following our ideas of
soldering \cite{ABW}, the two Lagrangeans were merged to yield a master
Lagrangean ${\cal L}_m={\cal L}_+\oplus {\cal L}_-$. The master action, in
any dimensions, was manifestly Lorentz or general coordinate invariant
and was also duality symmetric under both the groups mentioned
above. Moreover the process of soldering lifted the discrete group $Z_2
$ to its continuous version. The duality group for the master action
in either dimensionality therefore simplified to,
\be
\label{cm}
{\cal G}= O(2)\,\,\,;\,\,\, D=2k+2
\ee
Thus, at the level of the master action, the fundamental distinction 
between the odd and even $N$-forms gets obiliterated. It ought to be
stated that the  lack of usual Chern Simons terms in $D=4k+2$ dimensions
to act as the generators of duality transformations is compensated by the
presence of a similar term in the internal space. Thanks to this it was
possible to  explicitly construct the symmetry generators for the master
action in either two or four dimensions.

We also showed that the master actions in any dimensions, apart from
being duality symmetric under the $O(2)$ group, were factored, modulo a
normalisation,  as a
product of the self and anti self dual solutions,
\be
\label{1}
{\cal L}=\Big (F^\alpha +\tilde F^\alpha\Big) 
\Big (F^\alpha -\tilde F^\alpha\Big)\,\,\,;\,\,\,D=2k+2
\ee
where the internal index has been explicitly written and the
generalised Hodge operation was defined distinctly in $4k$ and
$4k+2$ dimensions. The key ingredient
in this construction was 
to provide a general definition of self duality $(\tilde{\tilde F}=F)$
that was applicable
for either odd or even $N$ forms. Self duality was now defined to
include the internal space and was implemented either by the
$\sigma_1$ or the $\epsilon$, depending on the dimensionality. 
This naturally led to the universal structure (\ref{1}).

Some other aspects of the analysis deserve attention. Specifically, the  
novel  duality symmetric actions obtained in two dimensions
revealed the interpolating role between duality and chirality.
Furthermore, certain points concerning the
interpretation of chirality symmetric action as the
analogue of the duality symmetric electromagnetic action in four
dimensions were clarified.
We also recall that the
soldering of actions to obtain a master action was an intrinsically
quantum phenomenon that
could be expressed in terms of an identity relating two ``gauge variant"
actions to a ``gauge invariant" form. The gauge invariance is with regard
to the set of transformations induced for effecting the soldering and has
nothing to  do with the conventional gauge transformations. In fact the
important thing is that the distinct actions must possess the self and
anti self dual aspects of some symmetry which are being soldered. The
identities obtained in this way are effectively a
generalisation of the usual Polyakov Weigmann identity. We conclude by
stressing
the practical nature of our approach to either bosonisation or duality
which can be extended to
other theories.

\bigskip

{\bf{Appendix: The Scalar Theory in 1+1 Dimensions}}

\bigskip

Here we discuss the effects of soldering in duality symmetry in
$1+1$ dimensions illuminating the similarities and distinctions
from the case of $3+1$ dimensions. 
It is simple to realise that the scalar theory is a very
natural example in these dimensions. For instance, there is no photon and
the Maxwell theory trivialises so that the electromagnetic field can be
identified with a scalar field. Thus all the results presented here can be
regarded as equally valid for the ``photon" field. 
Indeed the computations will also be presented in a very suggestive
notation which  illuminates the Maxwellian nature of the problem.

The Lagrangean for the free massless scalar field is given by,
\be
\label{w10}
{\cal L}=\frac{1}{2} \Big(\partial_\mu\phi\Big)^2
\ee
and the equation of motion reads,
\be
\label{w20}
\ddot\phi-\phi ''=0
\ee
where the dot and the prime denote derivatives with respect to time and
space components, respectively. Introduce a 
change of variables using electromagnetic symbols,
\be
E=\dot\phi\,\,\,\,;\,\,\,\, B=\phi '
\label{w30}
\ee
Obviously, $E$ and $B$ are not independent but constrained by the
identity, 
\be
\label{w40}
E'-\dot B=0
\ee
In these variables the equation of motion and the Lagrangean are expressed
as,
\br
\label{w50}
&\mbox{}&\dot E-B'=0\nonumber\\
&\mbox{}&{\cal L}=\frac{1}{2}\Big(E^2-B^2\Big)
\er
It is now easy to observe that the transformations,
\be
\label{w60}
E\rightarrow \pm B\,\,\,\,;\,\,\,\, B\rightarrow \pm E
\ee
display a duality between the equation of motion and the `Bianchi'-like
identity (\ref{w40}) but the Lagrangean changes its signature.
Note that there is a relative change in the signatures
of the duality transformations (\ref{m40}) and (\ref{w60}), arising
basically from dimensional considerations. This symmetry coresponds to the
improper group of rotations.

To illuminate the close connection with the Maxwell formulation, we
introduce covariant and contravariant vectors with a Minkowskian metric
$g_{00}=-g_{11}=1$, 
\be
F_\mu=\partial_\mu\phi\,\,\,;\,\,\, F^\mu=\partial^\mu\phi
\label{w60a}
\ee
whose components are just the `electric' and
`magnetic' fields defined earlier,
\be
F_\mu=\Big(E, B\Big)\,\,\,\,;\,\,\,\, F^\mu=\Big(E,- B\Big)
\label{w60b}
\ee
Likewise, with the convention $\epsilon_{01}=1$, the dual field is defined,
\br
\mbox{}^* F_\mu&=&\epsilon_{\mu\nu}\partial^\nu\phi =\epsilon_{\mu\nu}F^\nu 
\nonumber\\
&=& \Big(-B, -E\Big)
\label{w60c}
\er
The equations of motion and the `Bianchi' identity are now expressed by
typical electrodynamical relations,
\br
\partial_\mu F^\mu&=&0\nonumber\\
\partial_\mu \mbox{}^* F^\mu&=&0
\label{w60d}
\er

To expose a
Lagrangean duality symmetry, the basic principle of our approach 
to convert the original second order form
(\ref{w50}) to its first order version and then invoke a relabelling of
variables to provide an internal index, is adopted. This is easily achieved
by first introducing an auxiliary field,
\be
\label{w70}
{\cal L}= PE-\frac{1}{2}P^2-\frac{1}{2}B^2
\ee
where $E$ and $B$ have already been defined. The following renaming of
variables corresponding to the proper and improper transformations (see
for instance (\ref{matrix1}) and (\ref{matrix}) ) is used,
\br
\label{w80}
\phi&\rightarrow&\phi_1\nonumber\\
P&\rightarrow&\pm\phi_2'
\er
where we are just considering the discrete sets of the full
symmetry (\ref{matrix1}) and (\ref{matrix}). 
Then it is possible to recast (\ref{w70}) in the form,
\br
\label{w90} 
{\cal L}\rightarrow{\cal L}_\pm&=&\frac{1}{2}\Bigg[\pm
{\phi'}_\alpha\sigma_1^{\alpha\beta}\dot\phi_\beta -{\phi'}_\alpha^2\Bigg] 
\nonumber\\
&=&\frac{1}{2}\Bigg[\pm
B_\alpha\sigma_1^{\alpha\beta}E_\beta-B_\alpha^2\Bigg] 
\er
In the second line the
Lagrangean is expressed in terms of the electromagnetic variables.
This Lagrangean is
duality symmetric under the transformations of the basic scalar fields,
\be
\label{w100}
\phi_\alpha\rightarrow\sigma^{\alpha\beta}_1\phi_\beta
\ee
which, in the notation of $E$ and $B$, is given by,
\br
\label{w110}
B_\alpha &\rightarrow &\sigma^{\alpha\beta}_1B_\beta\nonumber\\
E_\alpha &\rightarrow &\sigma^{\alpha\beta}_1E_\beta
\er
It is quite interesting to observe that, contrary to 
the electromagnetic duality in $3+1$ dimensions,
the transformation matrix in the $O(2)$ space 
is a Pauli matrix, which is the discretised version of {\it
improper} rotations.
This result is in  agreement with that found from general algebraic
arguments \cite{SS, DGHT} which stated that for $d=4k+2$ dimensions
there is a discrete $\sigma_1$ symmetry. Observe that 
(\ref{w110}) is a manifestation of the
original duality (\ref{w60}) which was also effected by the same operation.
It is important to stress that the above symmetry is only
implementable at the discrete level. Moreover, since it is not connected
to the identity, there is no generator for this transformation.

To complete the picture, we also
mention that the following (proper) rotation,
\be
\phi_\alpha\rightarrow \epsilon_{\alpha\beta}\phi_\beta
\label{w120}
\ee
interchanges the Lagrangeans (\ref{w90}),
\be
{\cal L}_+\leftrightarrow {\cal L}_-
\label{w130}
\ee
Thus, except for a rearrangement of the the matrices generating the
various transformations, most features of
the electromagnetic example are perfectly retained. The crucial
point of departure is that now all these transformations are only discrete.
Interestingly, the master action constructed below lifts these symmetries
from the discrete to the continuous.

Let us therefore solder the two distinct Lagrangeans to
manifestly display the complete symmetries. Before doing this it is
instructive to unravel the self and anti-self dual aspects of these
Lagrangeans, which are essential to physically understand the soldering
process.  The equations of motion following from (\ref{w90}),  
in the language of the basic fields, are given by,
\be
\partial_\mu\phi_\alpha=\mp\sigma^1_{\alpha\beta}\epsilon_{\mu\nu}
\partial^\nu\phi_\beta
\label{dual1}
\ee
provided reasonable boundary conditions are assumed. Note that although
the duality symmetric Lagrangean is not manifestly Lorentz covariant, the
equations of motion possess this property. 
In terms of a vector field $F_\mu^\alpha$ and its dual $\mbox{}^*
F_\mu^\alpha$ defined analogously to
(\ref{w60b}), (\ref{w60c}), the equation of motion is rewritten as,
\be
\label{motion}
F_\mu^\alpha=\pm\sigma_1^{\alpha\beta}\mbox{}^* F_\mu^\beta=\pm\tilde
F_\mu^\alpha
\ee
where the generalised Hodge dual $(\tilde F)$ has been introduced.
Note the difference in the definition of this dual when compared
with the corresponding definition (\ref{m180}) in $3+1$ dimensions which
employs the epsilon matrix.
This explicitly reveals the self and anti-self dual nature of the
solutions in the combined internal and coordinate spaces. The result
can be extended to any $D=4k+2$ dimensions with suitable insertion of indices.

We now  solder the two Lagrangeans. 
This is best done by using the notation of the basic fields  
of the scalar theory. These Lagrangeans ${\cal L}_+$ and ${\cal L}_-$ 
are regarded as functions of the
independent scalar fields $\phi_\alpha$ and $\rho_\alpha$. 
Consider the gauging of the following symmetry,
\be
\delta\phi_\alpha =\delta\rho_\alpha=\eta_\alpha
\label{w140}
\ee
Following our iterative procedure
the final  Lagrangean  is obtained,
\be
{\cal L}={\cal L}_+(\phi)+{\cal
L}_-(\rho)-B_\alpha\Big(J_\alpha^+(\phi) +J_\alpha^-(\rho)\Big)-B_\alpha^2
\label{w150}
\ee
where the currents are given by,
\be
J_\alpha^\pm(\theta)=\pm\sigma_{\alpha\beta}^1\dot\theta_\beta-{\theta'}_\alpha
\,\,\,;\,\,\,\theta=\phi\,\,,\,\,\rho
\label{w160}
\ee
The above Lagrangean is gauge invariant under the extended
transformations including (\ref{w140}) and,
\be
\delta B_\alpha =\eta_{\alpha}'
\label{w170}
\ee
Eliminating the auxiliary $B_\alpha$ field using the equations of motion,
the final soldered Lagrangean is obtained from (\ref{w150}),
\be
{\cal L}(\Phi)= \frac{1}{4}\partial_\mu\Phi_\alpha \partial^\mu\Phi_\alpha
\label{w180}
\ee
where, expectedly, this is now only a function of the gauge invariant
variable, 
\be
\Phi_\alpha=\phi_\alpha-\rho_\alpha
\label{w190}
\ee
This master Lagrangean possesses all the symmetries
that are expressed by the continuous
transformations, 
\be
\label{205}
\Phi_\alpha\rightarrow R^\pm_{\alpha\beta}(\theta)\Phi_\beta
\ee
The generator corresponding to the $SO(2)$ transformations is easily obtained,
\br
\label{206}
Q&=&\int dy \Phi_\alpha \epsilon_{\alpha\beta} \Pi_\beta\nonumber\\
\Phi_\alpha\rightarrow \Phi'_\alpha &=&e^{-i\theta Q} \Phi_\alpha e^{i\theta
Q}
\er
where $\Pi_\alpha$ is the momentum conjugate to $\Phi_\alpha$.
Observe that either the original symmetry in $\sigma_1$ or the swapping
transformations were only at the discrete
level. The process of soldering has lifted these
transformations from the discrete to the continuous form. It is equally
important to reemphasize that the master action now possesses the $SO(2)$
symmetry which is more commonly associated with four dimensional duality
symmetric actions, and not for two dimensional theories. Note that by
using the electromagnetic symbols, the Lagrangean can be 
displayed in a form which manifests the soldering effect of
the self and anti self dual symmetries (\ref{motion}),
\be
\label{207}
{\cal L}=\frac{1}{8}\Big(F_\mu^\alpha+\tilde F_\mu^\alpha\Big)
\Big(F^\mu_\alpha-\tilde F^\mu_\alpha\Big)
\ee
where the generalised Hodge dual has been used.

An interesting observation is now made. Recall that the original duality
transformation (\ref{w60}) switching equations of motion into Bianchi
identities may be rephrased in the internal space by,
\br
\label{bianchi}
E_\alpha &\rightarrow& \mp R^\pm_{\alpha\beta} B_\beta\nonumber\\
B_\alpha &\rightarrow& \mp R^\pm_{\alpha\beta} E_\beta
\er
which is further written directly in terms of the scalar fields,
\be
\label{bianchi1}
\partial_\mu \Phi_\alpha \rightarrow \pm R^\pm_{\alpha\beta}
\epsilon_{\mu \nu}\partial^\nu \Phi_\beta
\ee
It is simple to verify that under these transformations even 
the Hamiltonian for the theories
described by the Lagreangeans
${\cal L}_\pm$ (\ref{w90}) are not invariant. However the Hamiltonian
following from the master
Lagrangean (\ref{w180}) preserves this symmetry. The Lagrangean itself
changes 
its signature. This is the exact analogue of the original situation.
A similar phenomenon also
occurred in the electromagnetic theory. 

It is now straightforward to give a Polyakov-Weigman type identity,
that relates the ``gauge invariant" Lagrangean with the non gauge invariant
structures, by reformulating (\ref{w180}) after a scaling of the fields
$(\phi, \rho)\rightarrow \sqrt 2(\phi, \rho)$,
\be
\label{w210}
{\cal L}(\Phi)= {\cal L}(\phi)+{\cal L}(\rho)-2\partial_+\phi_\alpha
\partial_-\rho_\alpha
\ee
where the light cone variables are defined in (\ref{35}).

Observe that, as in the electromagnetic example, the gauge invariance
is with regard to the transformations introduced for the soldering of the
symmetries. Thus, even if the theory does not have a gauge symmetry in the
usual sense, the dual symmetries of the theory can simulate the effects of
the former. This leads to a Polyakov-Wiegman type identity which has a
similar structure to the conventional identity.

It may be useful to highlight some other
aspects of duality and soldering 
which are peculiar to two dimensions, as for instance,
the chiral symmetry. The interpretation of this symmetry with regard to
duality seems, at least to us,
to be a source of some confusion and controversy. As is well known a
scalar field in two dimensions can be decomposed into two chiral pieces,
described by Floreanini Jackiw (FJ) actions \cite{FJ}. 
These actions are sometimes
regarded \cite{G} as the two dimensional analogues of the duality symmetric 
four dimensional
electromagnetic actions \cite{SS}. Such an interpretation is debatable
since the latter have the $SO(2)$ symmetry (characterised by an internal
index $\alpha$) which is obviously lacking in
the FJ actions. Our analysis, on the other hand, has shown how to
incorporate this symmetry in the two dimensional case. Hence we consider
the actions defined by (\ref{w90}) to be the true analogue of the duality
symmetric electromagnetic actions discussed earlier.
Moreover, by solving the equations
of motion of the FJ action, it is not possible to recover the second order
free scalar Lagrangean, quite in contrast to the electromagnetic theory 
\cite{SS}.
Nevertheless, since the FJ actions are just the chiral components of the
usual scalar action, these must be soldered to reproduce this result.
But if soldering is possible, such actions must also display the self and
anti-self dual aspects of chiral symmetry. This phenomenon is now explored
along with the soldering process.

The two FJ actions defined in terms of the independent scalar fields
$\phi_+$ and $\phi_-$ are given by,
\be
\label{w220}
{\cal L}^{FJ}_\pm(\phi_\pm)=\pm\dot\phi_\pm\phi_\pm'-\phi_\pm'\phi_\pm'
\ee
whose equations of motion show the self and anti self dual aspects,
\be
\label{w230}
\partial_\mu\phi_\pm=\mp\epsilon_{\mu\nu}\partial^\nu\phi_\pm
\ee
A trivial application of the soldering mechanism leads to,
\br
{\cal L}(\Phi)&=&{\cal L}^{FJ}_+(\phi_+)+{\cal L}^{FJ}_-(\phi_-)+\frac{1}{8}
\Big(J_+(\phi_+)+ J_-(\phi_-)\Big)^2\nonumber\\
&=&\frac{1}{2}\partial_\mu\Phi\partial^\mu\Phi
\label{w240}
\er
where the currents $J_\pm$ and the composite field $\Phi$ are given by,
\br
J_\pm&=&2\Big(\pm\dot\phi_\pm-\phi_\pm'\Big)\nonumber\\
\Phi &=& \phi_+-\phi_-
\label{w250}
\er
Thus the usual scalar action is obtained in terms of the composite field. 
The previous analysis has, however, shown that each of the Lagrangeans
(\ref{w90}) are equivalent to the usual scalar theory. Hence these
Lagrangeans contain both chiralities desribed by the FJ actions
(\ref{w220}). However, in the internal space, ${\cal L}_\pm$ carry the
self and anti self dual solutions, respectively. This clearly illuminates
the ubiquitous role of chirality versus duality in the two dimensional
theories which has been missed in the literature simply because, following
conventional analysis in four dimensions \cite{DT, SS}, 
only one particular duality symmetric Lagrangean ${\cal L}_-$
was imagined to exist. 

\bigskip

{\it Coupling to gravity}

\bigskip

It is easy to extend the analysis to include gravity. This is most
economically done by using the language of electrodynamics already
introduced. The Lagrangean for the scalar field coupled to gravity is
given by,
\be
{\cal L}= \frac{1}{2}\sqrt{-g}g^{\mu\nu}F_\mu F_\nu
\label{w260}
\ee
where $F_\mu$ is defined in (\ref{w60b}) and $g=\det g_{\mu\nu}$. Converting
the Lagrangean to its first order form, we obtain,
\be
\label{w270}
{\cal L}=P E
-\frac{1}{2\sqrt{-g}g^{00}}\Big(P^2+B^2\Big)+\frac{g^{01}}{g^{00}} P B
\ee
where the $E$ and $B$ fields are defined in (\ref{w30}) and $P$ is an
auxiliary field. Let us next invoke a change of variables mapping
$(E, B)\rightarrow (E_1, B_1)$ by means of the
$O(2)$ transformation analogous to (\ref{w80}),
and relabel the variable $P$ by $\pm B_2$.
Then the Lagrangean (\ref{w270}) assumes the distinct forms,
\be
\label{w280}
{\cal L}_\pm=
\frac{1}{2}\Bigg[\pm B_\alpha\sigma_{\alpha\beta}^1 E_\beta-\frac{1}
{\sqrt{-g}g^{00}} B_\alpha^2\pm \frac{g^{01}}{g^{00}}
\sigma_{\alpha\beta}^1 B_\alpha B_\beta\Bigg]
\ee
which are duality symmetric under the transformations (\ref{w110}). 
As in the flat metric, there is a swapping between ${\cal L}_+$
and ${\cal L}_-$ if the transformation matrix is $\epsilon_{\alpha\beta}$.
To obtain a duality symmetric action for all 
transformations it is necessary to construct the master action
obtained by soldering the two independent pieces. The dual aspects of the
symmetry that will be soldered are revealed by looking at the equations of
motion following from (\ref{w280}),
\be
\label{w290}
\sqrt{-g}F_\mu^\alpha=\mp g_{\mu\nu}\sigma_1^{\alpha\beta}\mbox{}^*
F^{\nu, \beta} \ee
The result of the soldering process, 
following from our standard techniques, leads to the master Lagrangean,
\be
\label{w310}
{\cal L}=\frac{1}{4}\sqrt{-g}g^{\mu\nu}F_\mu^\alpha F_\nu^\alpha
\ee
where $F_\mu^\alpha$ is defined in terms of the composite field given in
(\ref{w190}).
In the flat space this just reduces to the expression found previously in
(\ref{w180}). It may be pointed out that, starting from this master
action it is possible, by passing to a first order form, to recover the
original pieces. 

To conclude, we show how the FJ action now follows trivially by taking any
one particular form of the two Lagrangeans, say ${\cal L}_+$. To make
contact with the conventional expressions quoted in the literature
\cite{SO}, it is
useful to revert to the scalar field notation, so that,
\be
{\cal L}_+= \frac{1}{2}\Bigg[\phi_1'\dot\phi_2+\phi_2'\dot\phi_1
+2\frac{g^{01}}{g^{00}}\phi_1'\phi_2'-\frac{1}{g^{00}\sqrt{-g}}\phi_\alpha'
\phi_\alpha'\Bigg]
\label{w320}
\ee
This is diagonalised by the following choice of variables,
\br
\phi_1&=&\phi_+ +\phi_-\nonumber\\
\phi_2 &=&\phi_+ - \phi_-
\label{w330}
\er
leading to,
\be
\label{w340}
{\cal L}_+={\cal L}_+^{(+)} (\phi_+, {\cal G}_+)
+{\cal L}_+^{(-)} (\phi_-, {\cal G}_-)
\ee
with,
\br
\label{w350}
{\cal L}_+^{(\pm)} (\phi_\pm, {\cal G}_\pm)&=&\pm\dot\phi_\pm\phi_\pm'
+{\cal G}_\pm\phi_\pm'\phi_\pm'\nonumber\\
{\cal G}_\pm &=&\frac{1}{g^{00}}\Bigg(-\frac{1}{\sqrt{-g}}\pm g^{01}\Bigg)
\er
These are the usual FJ actions in curved space as given in \cite{SO}. Such
a structure was suggested by gauging the conformal symmetry of the free
scalar field and then confirmed by checking the classical invariance under
gauge and affine transformations \cite{SO}. Here we have derived this
result directly from the action of the scalar field minimally coupled to
gravity.


\newpage
\begin{thebibliography}{30}
\bibitem{S} M. Stone, University of Illinois Preprint, ILL-TH-28-89.
\bibitem{W} R. Amorim, A. Das and C. Wotzasek, Phys. Rev. D53 (1996) 5810.
\bibitem{ABW} E.M.C. Abreu, R. Banerjee and C. Wotzasek, Nucl. Phys. B509
(1998) 519.
\bibitem{AAR} For a review see, E. Abdalla, M.C.B. Abdalla and K.D. Rothe, 
{\it Nonperturbative
Methods in Two Dimensional Quantum Field Theory}, World Scientific,
Singapore, 1991.
\bibitem{M} E.C. Marino, Phys. Lett. B263 (1991) 63.
\bibitem{C} C. Burgess, C. L\"utken and F. Quevedo, Phys. Lett. B336 (1994)
18; E. Fradkin and F. Schaposnik, Phys. Lett. B338 (1994) 253; K. Ikegami,
K. Kondo and A. Nakamura, Prog. Theor. Phys. 95 (1996) 203; D.Barci, C.D.
Fosco and L. Oxman, Phys. Lett. B375 (1996) 267.
\bibitem{RB} R. Banerjee, Phys. Lett. B358 (1995) 297 and Nucl. Phys.
B465 (1996) 157.
\bibitem{RB1} R. Banerjee and E.C. Marino, hep-th/9607040 
Phys. Rev. D56 (1997) 3763 ; hep-th/9707100  Nucl. Phys. B507 (1997) 501.
\bibitem{AB} E. Abdalla and R. Banerjee, hep-th/9704176, Phys. Rev. Lett.,
80 (1998) 238.
\bibitem{RJ} See, for instance, R. Jackiw, {\it Diverse Topics in Theoretical
and Mathematical Physics}, World Scientific, Singapore, 1995.
\bibitem{RB2} R. Banerjee, Phys. Rev. Lett. 56 (1986) 1889.
\bibitem{RB3} N. Banerjee and R. Banerjee Nucl. Phys. B445 (1995) 516.
\bibitem{JR} R. Jackiw and R. Rajaraman, Phys. Rev. Lett. 54 (1985) 1219;
2060(E).
\bibitem{JS}J.Schwinger, Phys. Rev. 128 (1962) 2425.
\bibitem{AW} L. Alvarez-Gaume and E. Witten, Nucl.Phys.B234 (1983) 269.
\bibitem{SC} S. Coleman, Phys. Rev. D11 (1975) 2088.
\bibitem{DJT} S. Deser, R. Jackiw and S. Templeton, Ann. Phys. (NY)
140 (1982) 372; A.N. Redlich, Phys. Rev. D29 (1984) 2366.
\bibitem{BRR} R. Banerjee, H.J. Rothe and K.D. Rothe, Phys. Rev. D52
(1995) 3750.
\bibitem{TPN} P.K. Townsend, K. Pilch and P. van Nieuwenhuizen, Phys.
Lett. B136 (1984) 452.
\bibitem{DJ} S. Deser and R. Jackiw, Phys. Lett. B139 (1984) 371.
\bibitem{BR} R. Banerjee and H.J. Rothe, Nucl. Phys. B447 (1995) 183.
\bibitem {O} D.I. Olive, Exact electromagnetic duality, hep-th/9508089; Nucl.
Phys. B(Proc. Suppl.) 58 (1997) 43
\bibitem {Z} D. Zwanziger, Phys. Rev. D3 (1971) 880
\bibitem {DT} S. Deser and C. Teitelboim, Phys. Rev. D13 (1976) 1592
\bibitem {SS} J. Schwarz and A. Sen, Nucl. Phys. B411 (1994) 35
\bibitem {GR} G.W. Gibbons and D.A. Rasheed, Nucl. Phys. B454 (1995) 185
\bibitem {NB} N. Berkovits, Phys. Lett. B388 (1996) 743; ibid B398 (1997) 79
\bibitem {DGHT} S. Deser, A. Gomberoff, M. Henneaux and C. Teitelboim, Phys.
Lett. B400 (1997) 80
\bibitem {AG} L. Alvarez-Gaume and S.F. Hassan, Fortsch. Phys. 45 (1997) 159;
Also see, L. Alvarez-Gaume and F. Zamora, Duality in Quantum Field Theory
(String Theory), hep-th/9709180
\bibitem {KP} A. Khoudeir and N. Pantoja, Phys. Rev. D53 (1996) 5974
\bibitem {PST} P. Pasti, D. Sorokin and M. Tonin, Phys. Lett. B352 (1995) 59;
Phys. Rev. D52 (1995) R4277
\bibitem {G} H.O. Girotti, Phys. Rev. D55 (1997) 5136
\bibitem {FJ} R. Floreanini and R. Jackiw, Phys. Rev. Lett. 59 (1987) 1873
\bibitem {SO} J. Sonnenschein, Nucl. Phys. B309 (1988) 752
\end{thebibliography}
\end{document}